# Wavelets spectra of magnetization dynamics in geometry driven magnetic thin layers

Pawel Steblinski and Tomasz Blachowicz

Institute of Physics, Silesian University of Technology

Krzywoustego 2 str., 44-100 Gliwice, Poland

Squared cobalt thin layers of different thickness and width were investigated by numerical simulations. Using zero-valued externally applied magnetic field (geometry driven regime) and different initial conditions the magnetization dynamics were examined. The wavelet-based spectral analysis was applied. Transient states of different types were identified.

## I. Introduction

Nowadays nanotechnology techniques enable creation of low-dimensional magnetic single objects, or objects organized in structures, which can find challenging applications in magnetoelectronics. One of the main research tools, supporting above efforts, is the computer simulation with the use of realistic dimensions and material parameters. If the method bases on the Finite Element (FE) method, this give an opportunity to explore precisely influence of shape and edges onto the nano-devices performance. The widely-known software of this type is the Parallel Finite Element Micromagnetics Package (MAGPAR).[1]

In the current paper we describe results of simulations of squared, free, magnetic thin Co films. By employing continuous wavelets spectral analysis to describe magnetization dynamics, we identified different types of transient states, going on both in time and frequency domains.

## II. Wavelets for peak-positions detection

Wavelet analysis is an approach which enables simultaneous analysis of signals in frequency and time domains, thus deals with transients events which can be interpreted from both the frequency and time scales. Usually, the analysis is represented as a two-dimensional map which is constructed from an input signal. In Fig. 1. the idea of wavelet analysis is shown. The two time-events, which amplitudes are well located at moments $t_1$ and $t_2$, respectively, have the two frequencies $f_1$ and $f_2$, and finally were localized on the time-frequency map as the $W_1$, $W_2$, and $W_3$ events. In other words, at the same moment $t_1$, there are two physical events $W_1$ and $W_2$. On the other hand, the $f_1$ frequency is represented by the two events $W_1$ and $W_3$ separated in time.

In the current paper, in order to analyze positions of amplitudes (peak positions), the following, symmetrical Mexican-hat wavelet real-valued function is applied

$$g(z) = (z^2 - 1)e^{-z^2/2}, \qquad (1)$$

which can be then convoluted with the physical signal $f(t)$. The result is represented by the wavelet spectral decomposition map $W(\omega, T)$ (Fig. 1.) calculated as an integral for a given frequency $\omega$, taken from a range of frequencies, at the given time-moment $T$.[2] Thus, we have

$$W(\omega, T) = \sqrt{\omega} \int_{-\infty}^{+\infty} f(t) g[\omega(t - T)] dt. \qquad (2)$$

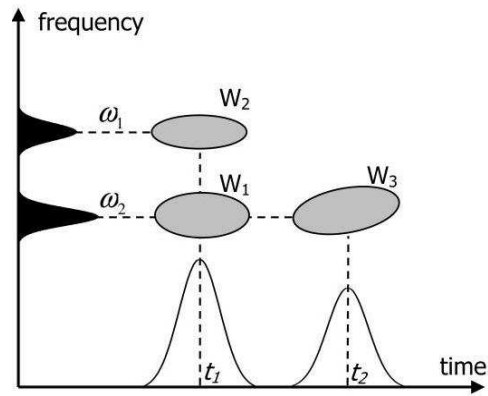

Fig. 1. The idea of 2-dimensional wavelet map $W(\omega, T)$. The two physical events $W_1$ and $W_2$ can be distinguished in the frequency scale, while the $\omega_2$ spectral component is composed from the two events, $W_1$ and $W_3$, separated in time at the moments $t_1$ and $t_2$.

In order to calculate the inverse wavelet transform (IWT) we can use the dual function $\Psi[\omega(t - T)]$ which satisfies the following condition of ortogonality

$$\int_0^{+\infty} d\omega \int_{-\infty}^{+\infty} |\omega|^3 g[\omega(t' - T)] \Psi[\omega(t - T)] dT = \delta(t - t') \qquad (3)$$

with the Dirac's delta function $\delta(t-t')$. Thus, we can express the original signal as

$$f(t) = \int_0^{+\infty} d\omega \int_{-\infty}^{+\infty} W(\omega,T)\,\omega\Psi[\omega(t-T)]dT. \qquad (4)$$

From that we can conclude, that wavelet spectral decomposition $W(\omega,T)$, seen qualitatively in Fig. 1., represents the amplitudes of dual basic functions of which the original signal is composed.

### III. Wavelets spectra of magnetization dynamics in thin magnetic layers

We perform simulations of magnetization evolution, and obtained spatio-temporal information, having applied assumed initial conditions for magnetization distribution. The simulated magnetization was geometry driven, thus, by demagnetizing fields depending on geometrical dimensions. The external field intensity was set to zero for all simulations performed.

The simulator employed the Landau-Lifshitz-Gilbert (LLG) equation in the following form

$$\frac{d\vec{M}}{dt} = -\frac{\gamma}{1+\alpha^2}\left(\vec{M}\times\vec{H}_{eff}\right) - \frac{\alpha}{M_s(1+\alpha^2)}\vec{M}\times\left[\vec{M}\times\vec{H}_{eff}\right], \qquad (13)$$

where $\vec{M}$ is the magnetization, $\gamma$ is the gyromagnetic factor, $\alpha$ is the dumping coefficient, $M_s$ is the magnetization at saturation, and where the effective magnetic field $\vec{H}_{eff}$ consists of the following several contributions

$$\vec{H}_{eff} = \vec{H}_{ani} + \vec{H}_{dem} + \vec{H}_{ex}. \qquad (14)$$

Thus, during simulations, we observed tendency of our system to evolve into the equilibrium via competition between the anisotropy field $\vec{H}_{ani}$, which is dependent on the crystalline structure of material, next, the demagnetization field $\vec{H}_{dem}$, which is dependent on a shape, and finally, the exchange field $\vec{H}_{ex}$ which includes contributions from quantum exchange interactions. Within the every simulation, we obtained the equilibrium state which satisfied the following condition

$$\vec{H}_{ani} + \vec{H}_{dem} + \vec{H}_{ex} = \vec{0}, \qquad (15)$$

while, for an arbitrary, transient state of evolution, we had

$$\vec{H}_{ani} + \vec{H}_{dem} + \vec{H}_{ex} \neq \vec{0}. \tag{16}$$

Using simulations we tested squared Co ultrathin layers of thickness ranging from 1.72nm up to 2.2nm, possessing edges dimensions between 25nm and 250nm. This range of edge-to-thickness ratios resulted in the in-plane states for magnetization at equilibrium.[3]

The simulations were realized for the following initial condition

$$M_x = 1, M_y = 0, M_z = 0, \tag{17}$$

For some simulated cases we obtained transient-oscillating states. However, for some cases we obtained simple evolutions with oscillating states excluded. The time-evolutions were obtained separately for the every $M_x, M_y, M_z$ magnetization component.

In order to perform time-resolved location of different frequency modes we used the wavelet spectral decomposition. For the sample of the 150nm width and the thickness equal to 2.2nm the interesting transient state for the $M_z$ component was obtained. The state disappeared after about 1.6 ns (Fig. 2). The different example of evolution of magnetization states was found for the $M_x$ component for the sample with width of 150 nm and thickness equal to 2.2 nm. The system evolved through the two transient wide states (Fig. 3). The another evolution for the $M_x$ component was obtained for the sample with edge length equal to 250nm and the thickness of 1.72nm (Fig. 4), where the magnetization evolved through several, subsequent oscillating states. There were obtained four types of transient regions: several increasing in frequency events (A), below 0.6ns, the wide doubled-in-time transient state at around 1.5ns (B), the single narrow case at 2.3ns (C), and finally, the frequency-doubled state visible at $t$=4.9ns (D).

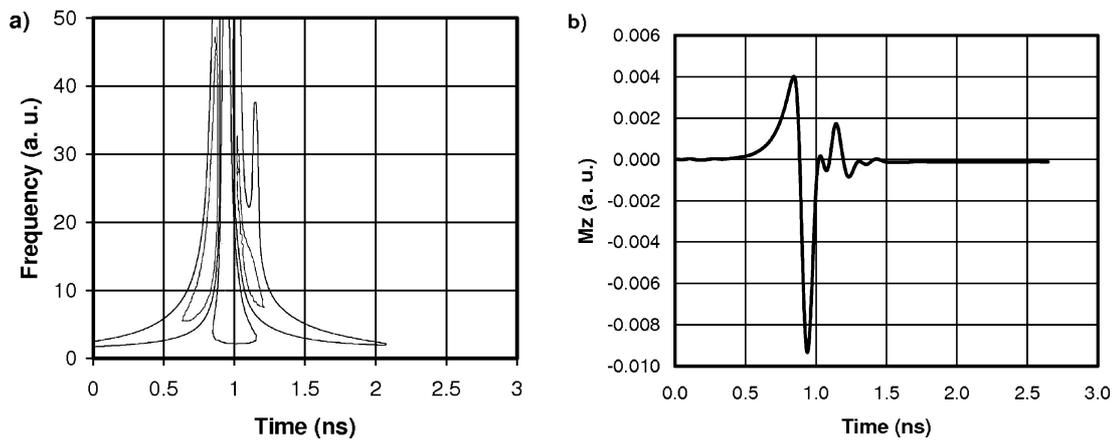

Fig. 2. The z-component magnetization dynamics for the $w=150$nm and $h=2.2$nm sample. The wavelet map (a), the time-evolution (b).

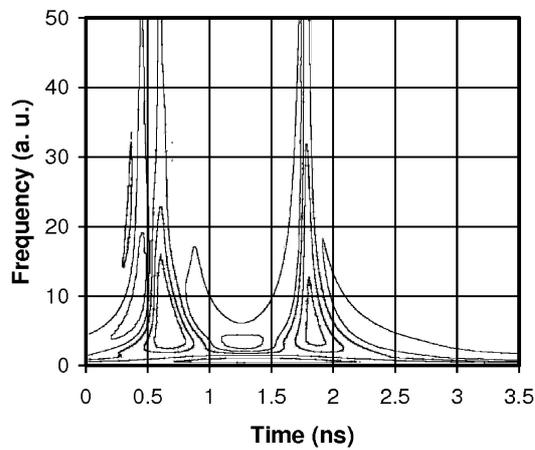

Fig. 3. Two transient events of the magnetization x-component for the $w=150$ nm and $h=2.2$ nm sample.

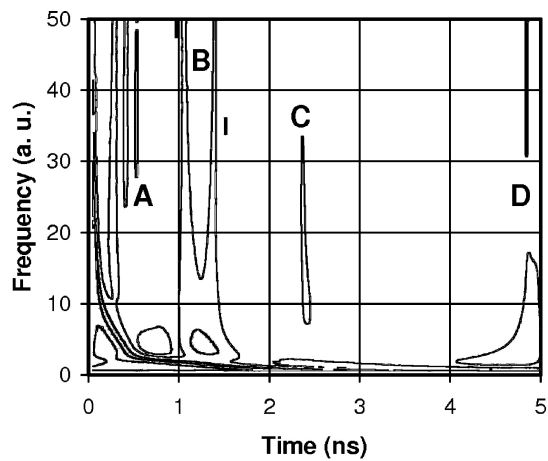

Fig. 4. The wavelet map of the magnetization x-component for the $w=250$nm and $h=1.72$nm sample.

## V. Conclusions

Within the current work we obtained different frequency-time behavior of magnetization in the low-dimensional squared cobalt nanodots. Thus, we obtained many different shapes of wavelet distributions of magnetization. From the signal theory point of view the wavelet approach is especially suitable for transient events as they take place in the cobalt nanoobjects we tested.

This type of spectral analysis can be additionally treated as a tool for vortex state investigations. This results from that fact, that an evolution leading to vortexes is associated with doubled-frequencies and higher order frequencies signals.[4-5] However, since the vortexes in some situations are not stable in time and space, then their behavior can be effectively described using wavelet analysis.